# INGENIERIE SYSTEME D'UN SYSTEME D'INFORMATION D'ENTREPRISE CENTRE SUR LE PRODUIT BASEE SUR UN CADRE DE MODELISATION MULTI-ECHELLES : APPLICATION A UN CAS D'ETUDE DE L'AIP LORRAIN.

**Jean-Philippe Auzelle (1), Olivier Nartz (2), Jean-Yves Bron (3)**
(1) Université Henri Poincaré-CRAN, Nancy, jean-philippe.auzelle@cran.uhp-nancy.fr
(2) Aip-Primeca Lorraine, Villers-lès-Nancy, olivier.nartz@aipl.uhp-nancy.fr
(3) Aip-Primeca Lorraine, Villers-lès-Nancy, jean-yves.bron@aipl.uhp-nancy.fr

**Résumé:**

*A travers ses projets, l'Atelier Inter-établissements de Productique Lorrain (AIPL), en tant que MOA (maître d'ouvrage) et MOE (maître d'œuvre) de rang 1, a la volonté de fournir à ses clients (enseignants, filières de formation, étudiants etc.) des supports didactiques crédibles, à l'échelle d'une infrastructure industrielle réelle agile de production de biens et de services. Dans ce contexte évolutif , sa Direction a choisi de faire disparaître le concept CIM qui proposait un pilotage d'entreprise intégrée, au profit d'un pilotage de systèmes d'informations (SI) distribués, hétérogènes, autonomes et évolutifs au gré des coopérations éphémères entre des partenaires industriels qui échangent dorénavant des flux informationnels et matériels. Ces aspects sont étudiés en recherche au CRAN dans le cadre d'une thèse qui s'appuie sur l'aspect récursif des systèmes et de leurs modélisations, ainsi que sur leurs aspects multi-échelles et multi-points de vues pour proposer une méthodologie d'ingénierie système basée sur les modèles (ISBM) d'un SI centré sur le produit. Afin de valider ces travaux de recherche, cette ISBM de la MOA et de la MOE de rang 1 a été mise en œuvre sur un cas d'étude à l'AIPL : celui du projet « eFormation en eProduction ».*

## 1 Introduction

Ce papier se veut être une présentation de la mise en œuvre, à l'Atelier Inter-établissements de Productique Lorrain[1], d'une ingénierie de son système d'information dans le cadre du projet « eFormation en eProduction » dont l'objectif est de proposer un support didactique suffisamment complexe et surtout crédible en comparaison à une structure industrielle de production de biens et de services. Les travaux cités dans cet article sont relatifs au sous-projet « traçabilité du produit » et plus particulièrement à la réception des matières premières et des composants. L'étude de cas menée est issue d'une collaboration entre l'équipe projet « Système Contrôlé par le Produit » du Groupe thématique SYMPA (Systèmes de production ambiants) au sein du CRAN (Centre de Recherche en Automatique de Nancy) et l' « Équipe de Recherche en processus Innovatifs » de L'École Nationale Supérieure en Génie des Systèmes Industriels (ENSGSI) de l'INPL (Institut National Polytechnique de Lorraine) [**1**].

[1] L'AIPL est l'un des deux sites géographiques composant le pôle AIP-PRIMECA Lorraine (www.aip-primeca.net/lorraine)

Les travaux s'inscrivent dans une démarche d'**Ingénierie Système** (IS) promue par l'AFIS[2] qui identifie [**2**] le « **Système À Faire** » comme étant le système attendu (le système cible) résultant d'un Système-projet « **pour faire** » guidé par les processus de bonnes pratiques de l'IS [**3**]. Dans cette étude, le processus proposé par les travaux de recherche se fonde sur une **Ingénierie Système Basée sur les Modèles** (ISBM) guidée par un cadre de modélisation multi-échelles.

Nous présentons au paragraphe 2 de cet article, le contexte de notre cas d'étude d'un point de vue industriel, d'un point de vue scientifique, puis d'un point de vue formation. Au paragraphe 3 est présenté le cadre de modélisation proposé par les travaux de thèse à la base de notre ingénierie. Ensuite, nous détaillons partiellement la mise en œuvre d'une telle ingénierie appliquée au sous-projet « traçabilité du produit » du projet « eFormation en eProduction » pour la réception matières à l'AIPL. Enfin, nous concluons cet article en exposant les retours d'expériences de la mise en œuvre de cette méthodologie à l'AIPL.

# 2 Contexte de l'étude de cas

## 2.1 Contexte et problématique industriel

### 2.1.1 Le Système-à-Faire

L'évolution du paradigme d'intégration vers celui d'interopération afin de satisfaire aux nouveaux challenges industriels [**4, 5**] pousse à résoudre des problématiques d'interopération [**6**] entre des éléments constitutifs (COTS) répartis dans les différents domaines fonctionnels du système d'entreprise. Ceux-ci fonctionnent ensemble et échangent des informations en vue de satisfaire les objectifs définis pour la mission qui leur est affectée par un donneur d'ordres et de répondre aux exigences stratégiques, techniques, réglementaires et légales. Ces problématiques se complexifient lorsque les entreprises souhaitent coopérer (à plus ou moins long terme) au sein d'un réseau réticulaire d'entreprises pour participer à une mission commune qui consiste à fournir des biens ou des services. Cette nouvelle vision, implique de considérer l'entreprise comme un Système-Entreprise (**Système À Faire**) qu'il soit une entreprise individuelle ou un réseau d'entreprises, afin de le concevoir comme un Système-Entreprise ouvert [**7**] pour une mission donnée.

### 2.1.2 Le Système-Projet

La conception d'un tel système ouvert impose que l'interopérabilité soit une de ses caractéristiques fonctionnelles afin de développer des SI distribués, hétérogènes, autonomes et en constante évolution, permettant à leurs éléments constitutifs d'établir une collaboration agile pilotée par la maîtrise d'ouvrage (MOA) du système bénéficiaire. Pour réaliser l'ingénierie d'un tel « Système À Faire » (SAF), la MOA s'assure de la cohérence des compétences métiers de la maîtrise d'œuvre (MOE) « pour faire » la spécification d'une part, des relations d'interopération entre les SI constitutifs, et d'autre part, des SI en tant que composants du SE, au travers d'un « Système Pour Faire » appelé aussi « Système-projet ». Cette étude se justifie aussi par le besoin de capitaliser des connaissances en termes de modélisation d'entreprise afin de garantir et d'internaliser un savoir-faire d'ingénierie propre à chaque entreprise ou réseaux d'entreprise [**8**]. Considéré comme un tout cohérent, il convient de prendre en compte les aspects multi-points de vue et multi-échelles induits du SAF, au travers de l'ensemble des activités d'une ingénierie permettant de passer de l'expression du besoin (préparation), à sa définition, jusqu'à sa spécification détaillée. La difficulté majeure rencontrée dans l'ingénierie de SI « **à-Faire** » distribués, hétérogènes et autonomes, concerne la complexité de leur assemblage pour former un SI d'entreprise cohérent au regard d'une performance globale à atteindre. En effet, les applications supports de ces SI étant pour la plupart considérées comme des composants sur étagères (COTS), leurs configurations doivent prendre en compte les propriétés et les fonctionnalités d'un assemblage plus global auquel elles vont participer tout en conservant les conditions opérationnelles propres à chacune. Par analogie, ces problématiques peuvent être assimilées à celles rencontrées dans les Systèmes de Systèmes (SdS) [**9, 10, 11**]. La structure même de ces SdS contraint à considérer l'ingénierie système sous un autre angle : celui d'une Ingénierie de SdS.

[2] Association Française d'Ingénierie Système (http://www.afis.fr)

### 2.2 Contexte et problématique scientifique

Le contexte scientifique est principalement celui sur lequel J.Ph. Auzelle s'est fondé dans le cadre de sa thèse, notamment au travers du concept de SdS « à-Faire » (le système-cible) et « pour-Faire » (le système-projet) et de leurs aspects récursifs.

#### 2.2.1 Le système-à-Faire : un Système de Systèmes d'Information (SdSI)

Le concept de SdS et plus particulièrement les critères fonctionnels qui le définissent, permettent de distinguer les aspects fonctionnels des aspects techniques d'un assemblage d'un SI distribué, hétérogène et autonome, afin d'optimiser plus facilement la configuration de cet assemblage et de favoriser son agilité pour faire face à un environnement économique changeant [**1**]. D'autre part, ces critères fonctionnels permettent de s'affranchir de l'obsolescence [**12**] des technologies de l'information, en remontant d'un COTS pouvant devenir obsolète, à sa fonction, voire à l'exigence concernée, pour mieux le cerner individuellement tout en permettant d'exercer une véritable gestion de sa configuration dans le but de favoriser ses interopérations au sein d'un **SdS d'Information** [**1**] (SdSI). Les problématiques liées à la reconfiguration ont orienté les travaux de recherche qui s'appuient notamment sur une interprétation du paradigme « Holonic Manufacturing Systems » (HSM) de l'initiative « Intelligent Manufacturing Systems » (IMS) en phase avec les travaux du projet Européen « FP6/PABADIS'PROMISE » [**13**]. Un des objectifs est d'expérimenter l'intérêt de rendre le produit et, au delà l'ensemble du procédé, interactif afin d'organiser de façon plus collaborative l'interopérationnalité des différents systèmes hétérogènes de pilotage et de gestion (APS, ERP, MES, CRM) [**14**] composant, au sens d'un SdS. Cet objectif est partagé par le CRAN au travers de son équipe projet « Système Contrôlé par le Produit » qui étudie les concepts mais aussi les aspects techniques de mise en œuvre pour *tirer parti des progrès et de la miniaturisation croissante des technologies infotroniques (RFID, communications sans-fils, etc.) et mécatroniques (composants logiciels embarqués) pour faire de ce produit actif un objet composite assurant une relation récursive logiciel-matériel entre les services et les biens associés aux produits en reliant tout objet logique de contrôle ou de gestion à au moins un objet physique du procédé »* [**5**].

#### 2.2.2 Le Système-Projet : l'ingénierie d'un SdSI

Les travaux de thèse sur laquelle nous nous appuyons, tiennent compte du contexte cité ci-dessus et se veulent une contribution à l'ingénierie d'un système d'information en proposant une **Ingénierie Système Basée sur les Modèles** (ISBM) [**15**] afin de favoriser les interopérations au sein de systèmes complexes à grande échelle que sont les entreprises étendues ou les réseaux d'entreprise centrés autour d'une finalité : le Produit. Basée sur un cadre de modélisation, cette ingénierie vient en appui des outils et méthodes d'Ingénierie Système (IS) déjà existants et son rôle est de fournir un cadre de modélisation structuré, récursif, multi-échelles, et multi-points de vue, de grands systèmes complexes, au travers d'invariants de modélisation. A partir d'un état de l'art sur les cadres de modélisation, Auzelle a retenu le **cadre de modélisation de Zachman** [**16, 17**] (CMZ) parce qu'il est adapté aux Systèmes d'Information, qu'il propose un guide simple et relativement générique, qu'il laisse une grande liberté d'instanciation à son utilisateur dans son action de modélisation, qu'il permet la spécification d'un système à des niveaux d'observations différents et qu'il permet de le faire de manière récursive. Pour préserver le savoir-faire relatif à son Système, le procédé d'Ingénierie Système Basée sur des Modèles (ISBM) [**18**] proposé dans la thèse de Auzelle, considère que l'Entreprise doit faire office de maîtrise d'ouvrage (MOA) et de maîtrise d'œuvre de rang 1 (MOE-1) « à minima » aux niveaux contextuels et conceptuels des cadres de modélisation de Zachman en assurant une relation syntaxique par héritage d'invariants de modélisation. Cela ne préjuge pas qu'un métier particulier, par exemple pour un COTS, n'utilise pas une Ingénierie Dirigée par des Modèles (IDM) [**19**] pour assurer la transformation des modèles (par exemple, en CFAO du Système-Produit) menant au déploiement de l'application à partir de l'invariant hérité (par exemple, un modèle sémantique de données en cellule « Conceptual-What » du cadre de modélisation de Zachman).

### 2.3 Contexte de formation

L'objectif du projet « eFormation en eProduction » est de former des étudiants[3] à l'ingénierie de systèmes d'information dans un contexte plausible de eProduction, fourni par l'AIPL dont le rôle est d'agir en tant que MOA et MOE de rang 1 au service de la pédagogie à mettre en place.

#### 2.3.1 Le Système-à-Faire : le système eProduction

Dans ce contexte, l'AIPL se doit de fournir le système à étudier. Comme le montre la *Figure 1*, le projet intègre de nombreux outils (ERP, MES, …) et donc autant de problématiques d'interopération entre ces éléments constitutifs (COTS).

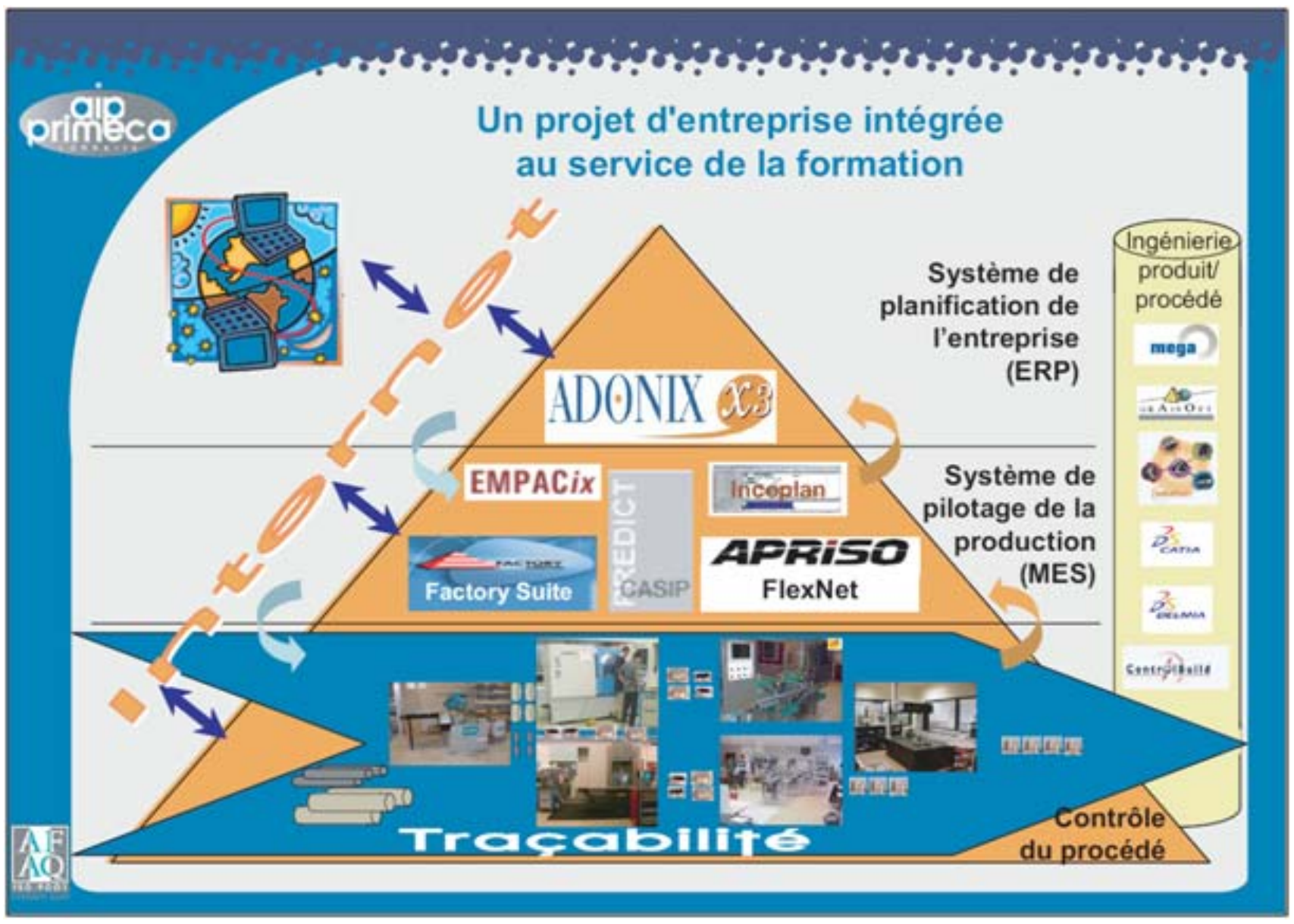

*Figure 1. Projet eProduction de l'AIP-Priméca Lorraine*

Néanmoins, l'enseignement de l'ingénierie de systèmes complexes reste très difficile. Ceci est en partie lié à l'absence de méthodologie relativement simple à appliquer dès lors que des projets d'étudiants mettent en activité plusieurs COTS. En effet, bien que le projet « eProduction » se doive d'être un projet fédérateur d'un ensemble de sous-projets liés à la formation des étudiants, on constate que les ingénieries déployées pour ces sous-projets restent parcellaires, isolées et manquent de cohérences entres-elles et ne répondent pas forcément aux exigences et objectifs du projet global.

#### 2.3.2 Le Système-Projet : application d'une IS pour le système eProduction

Il est donc nécessaire d'avoir une vue globale avec des objectifs et des exigences globaux bien identifiés, pour que chacun des sous-projets à constituer puisse les décliner et contribuer ainsi à atteindre les objectifs globaux tout en respectant les exigences imposées, de manière à garantir un système-AIPL cohérent. Des outils sont mis en place pour permettre une certaine forme de capitalisation des ingénieries déployées lors de projets au travers d'espaces de travail coopératif (QuickPlace). Ces espaces permettent aux acteurs de sous-projets d'échanger des documents et de conserver les traces de leur conception. L'ensemble des espaces de sous-projets sont stockés d'une année universitaire sur l'autre et permet leur ré-exploitation lorsque des projets s'étalent sur plusieurs années. Les étudiants et les enseignants peuvent ainsi réutiliser les ingénieries déployées lors de projets antérieurs et contribuer à les enrichir. D'autre part, le système de management de la Qualité mis en place par l'AIPL, qui lui permet d'être certifié ISO9001:2000 depuis 2005, participe à sa façon à capitaliser les savoir-faire métier au travers des modèles de processus internes qui sont formalisés à l'aide de l'atelier logiciel MEGA Modelling Suite[4].

[3] Nancy-Université : Master Ingénierie de Systèmes Complexes, ESIAL : École Supérieure d'Informatique et Application de Lorraine, ENSGSI : École Nationale Supérieure en Génie des Systèmes Industriels

L'étude menée en collaboration avec le CRAN vise à renforcer l'axe méthodologique d'ingénierie système, en cohérence avec l'implication du pôle AIP-Priméca Lorraine et du réseau national AIP-Priméca dans les travaux de l'AFIS. Elle s'inscrit clairement dans la stratégie du pôle pour répondre aux nouvelles demandes des filières de formation en ingénierie système avec la mise en place d'un projet fédérateur « Ingénierie numérique et systèmes ambiants ».

## 3 Recherche & développement en Ingénierie de SdSI

### 3.1 Le Système-à-Faire : une Ingénierie de SdSI

Compte tenu des aspects récursifs des systèmes constitutifs, il convient conjointement de rendre récursives leurs ingénieries au travers d'un CMZ particulier. Ainsi, cette approche récursive d'ingénierie peut contribuer à réduire la complexité de la modélisation de grands systèmes complexes sur des échelles différentes, selon des niveaux d'observation différents et avec des points de vue différents. J.Ph. Auzelle [**1**] propose que ce CMZ s'apparente à une sorte de « pattern » où chaque objet de modélisation est en lien avec un autre au travers de relations et de règles de modélisation. Ce sont ces relations et les règles de modélisation qui garantissent la cohérence entre les modèles distribués et capitalisés au sein de chaque CMZ mais aussi au sein de chacun des CMZ propres aux ingénieries particulières et distribuées des SI constitutifs. Avec ce mécanisme de récursivité de l'ingénierie basée sur des modèles cohérents et capitalisés au sein de multiples CMZ associés à chaque SAF et à chaque système constitutif, la MOA et la MOE vont pouvoir partager et capitaliser leurs modèles à travers le temps. La méthodologie proposée dans la thèse de J.Ph. Auzelle a principalement pour objectifs d'assurer la cohérence de la modélisation d'un assemblage de systèmes d'information appartenant à un réseau réticulaire de Systèmes-Entreprises, et de faciliter les interopérations entre leurs applications au gré des collaborations conjoncturelles. Cette cohérence est assurée par les relations qui lient les éléments de modélisation créés par la MOA dont l'objectif est de réaliser un Système à Faire que l'on qualifiera de SAF-Système-entreprise. De par l'assemblage récursif des éléments constitutifs de ce SAF-global, les différentes MOE participant à la réalisation de ce SAF-Système-entreprise vont hériter des éléments de modélisation et des modèles de la MOA pour la réalisation de leur propre SAF-Sn. La mise en œuvre de l'ingénierie qui participe à ces héritages et à la mise en cohérence des modèles de l'ensemble des ingénieries des SAF, au travers du cadre de modélisation de Zachman, est appelée l'Ingénierie Système Basée sur les Modèles (ISBM) [**1**].

### 3.2 Le Système-projet : application d'une ISBM outillée par un Prototype de modélisation multi-échelle basé sur le cadre de modélisation de Zachman

Avant la mise en œuvre d'une telle ISBM, la MOA doit se poser la question sur la réelle utilité d'une telle démarche pour son SAF. Cette ISBM est-elle adaptée à la problématique que doit résoudre la MOA ? Ainsi celle-ci doit se poser la question suivante : pourquoi modélise-t-on ?
Si le « Système à Faire » justifie la mise en œuvre d'une ISBM, la MOA doit définir le contexte dans lequel doit se dérouler la modélisation afin d'assurer son pilotage, à savoir :

- Qui modélise ? Attribution d'un rôle pour chacun des acteurs du système projet (le Système pour Faire) : MOA, MOE 1 & 2, etc.
- Où modéliser ? Attribution à chacun des acteurs d'un niveau d'observation dans le CMZ pour sa propre modélisation : niveaux « contextual – conceptual – logical – physical - technical ».
- Quoi modéliser ? Définition des tpes de modélisations que chaque acteur peut employer à chaque perspective d'observation du CMZ (« what – how – where – who – when – why ») : UML, SYSML, BPMN etc.
- Quand modéliser ? Organisation temporelle du système projet (le Système pour Faire) avec des jalons pour ponctuer les moments clés du projet : préparation, définition et spécification.

[4] http://www.mega.com

- Comment modéliser ? Description détaillée des enchaînements de modélisation pour chaque acteur du système projet.

Afin de supporter notre méthodologie de modélisation, les travaux de recherche ont permis de prototyper un outillage, sur la base d'une adaptation de l'environnement de modélisation MEGA Modelling Suite. En effet, cet environnement fournit un ensemble d'outils de modélisation structuré autour d'un méta-modèle commun paramétrable et une base de modèles unifiée assurant une cohérence inter-modèles. MEGA Modelling Suite offre aussi un portail de représentation de l'architecture des modèles d'un projet au travers du CMZ. Comme dans les travaux de thèse de Baïna [**20**], nous utiliserons ce portail comme point d'entrée de notre modélisation. Notre problématique de modélisation récursive, basée sur le CMZ, nous a amenés à étendre le méta-modèle unifié de l'environnement MEGA Modelling Suite afin de formaliser le concept de « point de vue » relatif au CMZ, ainsi que la spécification relative d'une typologie des systèmes (bénéficiaires et/ou constitutifs) au travers de leur syntaxe abstraite représentée sous la forme de paquetages. Ainsi, un système constitutif participe à la réalisation de la mission d'un SdSI, système bénéficiaire. Pour un système donné, représenté par son paquetage, est associée une instance de CMZ au travers d'un ensemble de cellules spécifiant un point de vue de ce système. Ce point de vue est caractérisé par son niveau d'abstraction (Contextual, Conceptual, Logical, Technical, Operational) et sa perspective d'observation (What, How, Where, Who, When, Why). La syntaxe abstraite de ce concept est formalisée par la méta-classe « Point De Vue » associée à un paquetage.

# 4 Cas d'étude

## 4.1 Le Système-à-Faire : le système de traçabilité du produit en réception matières

Le contexte général de notre étude (*Figure 2* – encadré « cas d'étude ») s'inscrit dans un scénario PLM global, constituant le cadre de recherche du groupe « systèmes interopérants » du CRAN (Centre de Recherche en Automatique de Nancy) en partenariat avec le département d'innovation mécanique et de gestion en Italie (DIMEG) [**21**].

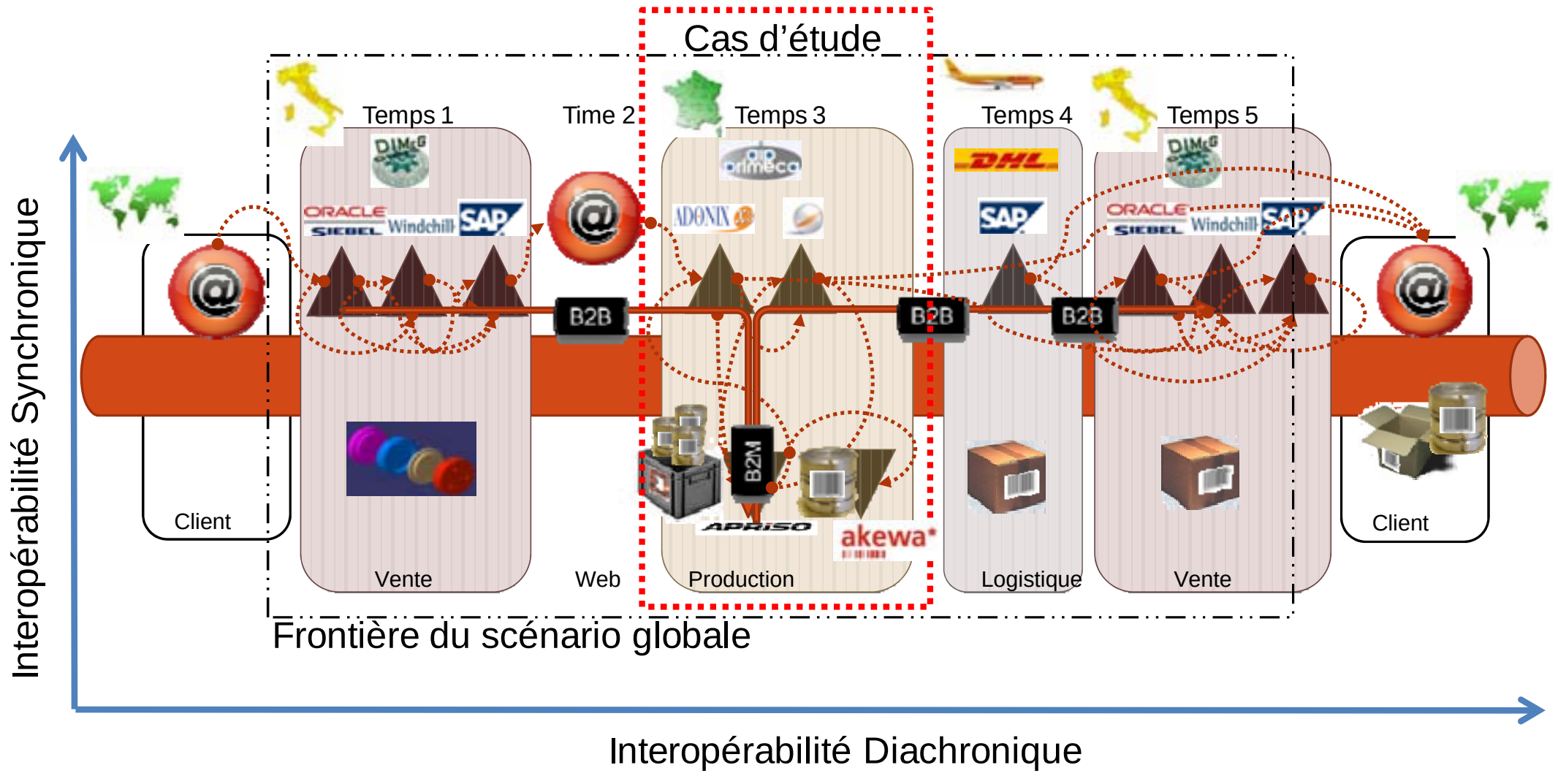

*Figure 2. Scénario Global*

Le contexte particulier de nos travaux concerne la production manufacturière dans un environnement multi-sites comprenant un site de conception (DIMEG en Italie), un site de production (AIPL en France) et un transporteur (DHL). Chaque site contribue au cycle de vie du produit et produit/consomme de l'information nécessaire à la réalisation de ses processus. A cette fin, chaque site échange des flux physiques et informationnels relatifs au produit à fournir au client final. Les sites sont distribués en Europe, sont autonomes opérationnellement et managérialement, ils évoluent au cours du temps et sont hétérogènes de part leurs propriétés intrinsèques.

Le cas d'étude traité dans nos travaux concerne l'ingénierie d'un **système de traçabilité des produits** sur le site de production. Pour illustrer nos propos, nous mettrons en œuvre la méthodologie de modélisation afin de spécifier rigoureusement le cahier des charges de développement de notre SAF (SAF-eProduction), avec une focale particulière sur l'interopérabilité entre les applications concernées par le processus de réception matières (le « SAF-Traçabilité_du_produit » Système constitutif du Système bénéficiaire « SAF-eProduction ») (Figure 3- zone encadrée). L'AIPL fournit une infrastructure de type industrielle comprenant des ingénieurs, des machines-outils, une ligne de montage automatique, un logiciel de conception (CAD), un ERP et un MES. DIMEG fournit une infrastructure pour la conception et la commercialisation de pièces mécaniques, comprenant des ingénieurs, un CAD, un ERP, un PDM, et un CRM.

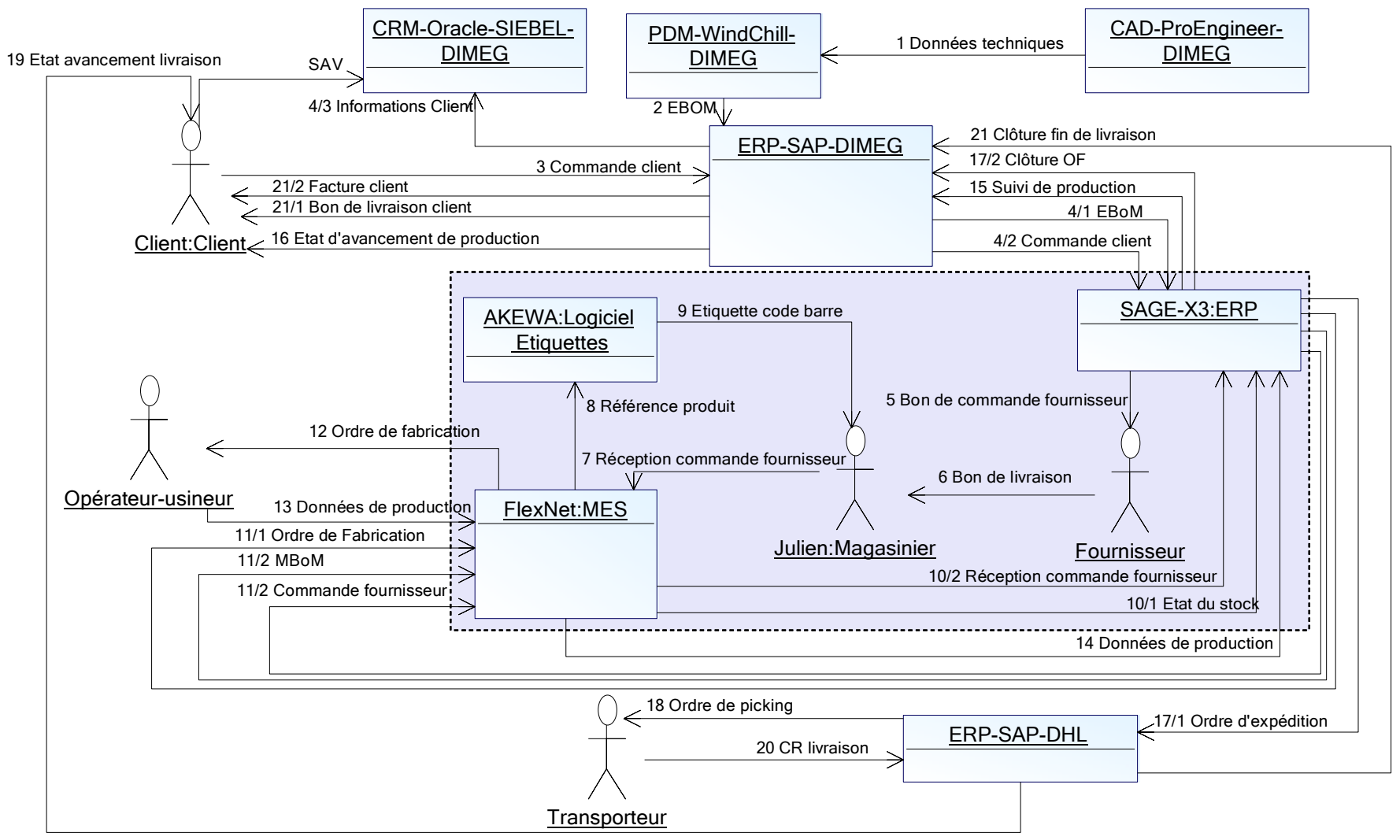

*Figure 3. Diagramme de collaboration partiel du scénario global*

### 4.2 Le Système-projet : application d'une ISBM pour le système cible

La première étape de notre ingénierie a consisté, en tant que MOA, à lire attentivement le cahier des charges proposé par le donneur d'ordres et à échanger avec celui-ci afin d'élaborer et de valider un premier diagramme d'Objectifs et d'Exigences (*Figure 4*) proposant une décomposition des objectifs principaux en sous-objectifs auxquels sont associées des exigences (réglementaires, techniques, choix d'entreprise etc.) qui peuvent elles-mêmes être décomposées en sous-exigences. Ce modèle a été placé dans la cellule intersection du niveau d'observation « Why » et du point de vue « Contextual » du CMZ de notre SAF-eProduction.

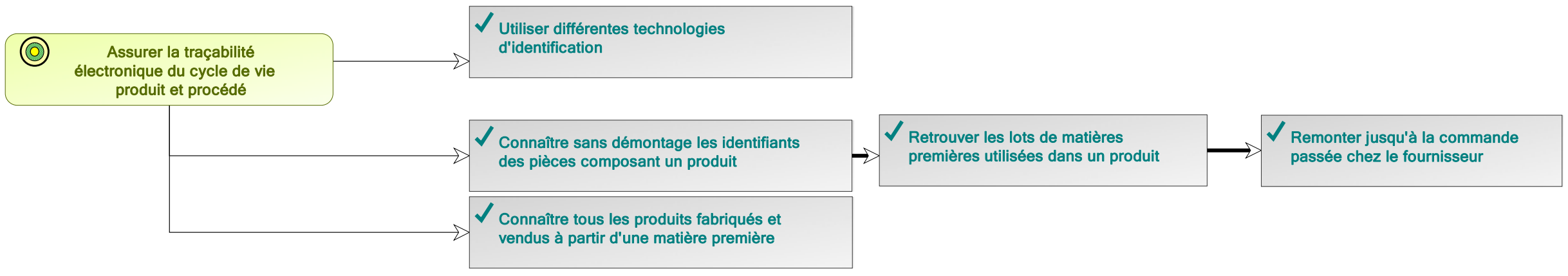

*Figure 4 : Diagramme d'Objectifs et d'Exigences (vue partielle)*

Nous avons créé parallèlement dans la cellule « What » du point de vue « Contextual », un diagramme de paquetage composé initialement du seul paquetage correspondant au SI du SAF de notre projet. Ce diagramme s'est enrichi par la suite des paquetages des Systèmes contributeurs pour former une vue d'ensemble des ingénieries des applications de notre projet.

Le domaine fonctionnel de notre SAF-eProduction étant celui de la production de biens et de services, nous avons choisi de dériver les processus du même domaine proposés par le modèle MES de MESA International [**22**]. La spécification des flux informationnels échangés entre ces processus (et leurs sous-processus) est dérivée du modèle fonctionnel proposé par le modèle PRM [**23**]. Nous avons spécifié ainsi une vue d'ensemble des processus de notre SAF et leurs interrelations informationnelles (*Figure 5*) que nous avons placés au niveau d'observation « How ». Cette vue a été enrichie par l'affectation des exigences issues du diagramme d'Objectifs et d'Exigences précédemment créé, aux processus eux-mêmes et par propagation à leurs sous-processus. Notre étude de cas détaillée étant relative à la réception matières, nous nous sommes focalisés sur le processus 4.3 (Gestion des réceptions matières premières et composants), sous processus du processus 4.0 (Gestion des matières premières et composants) de notre modèle et ses interactions avec les processus connexes (*Figure 6*).

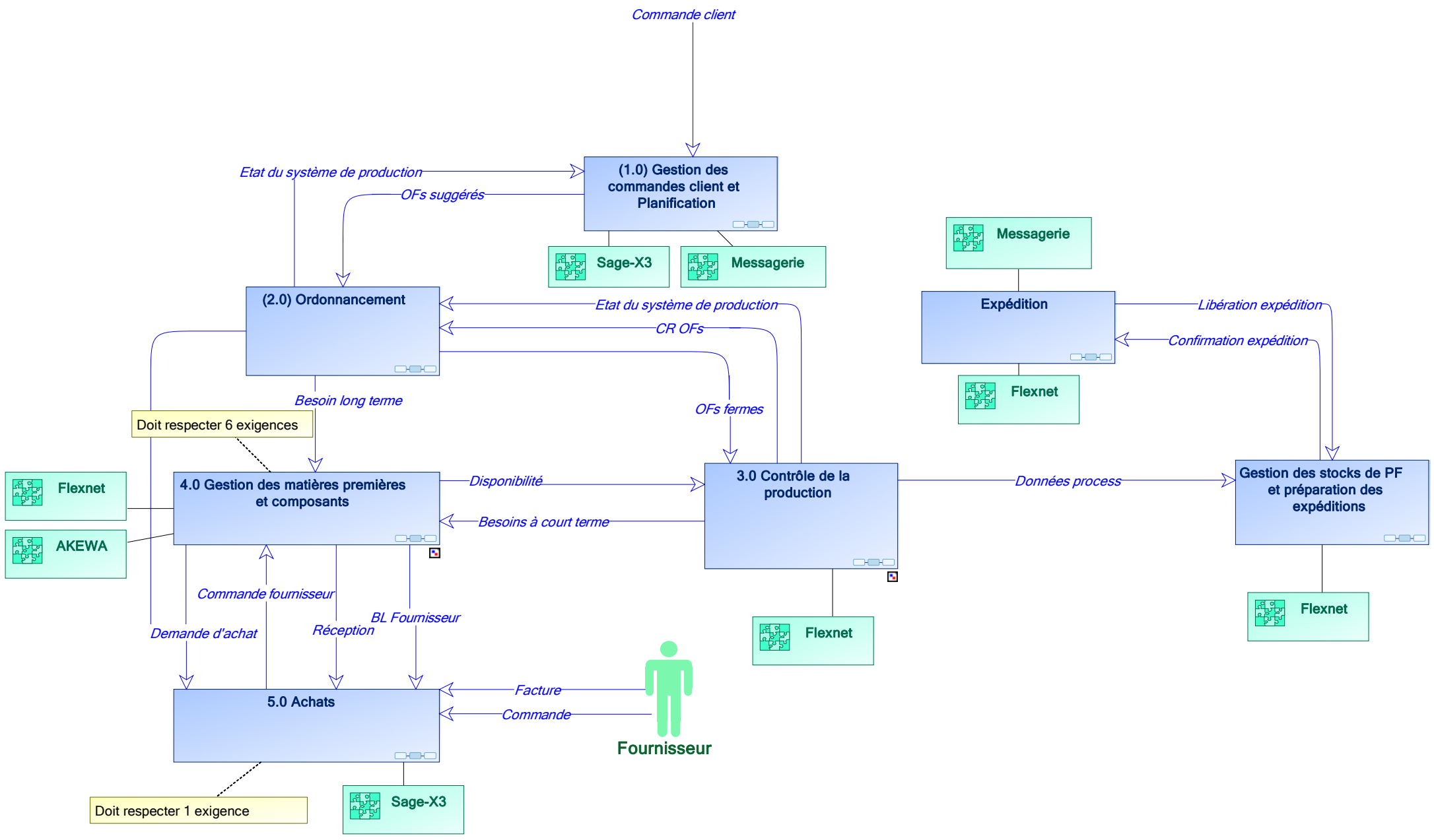

*Figure 5. Diagramme de processus du SAF « eProduction », dérivé du modèle fonctionnel PRM*

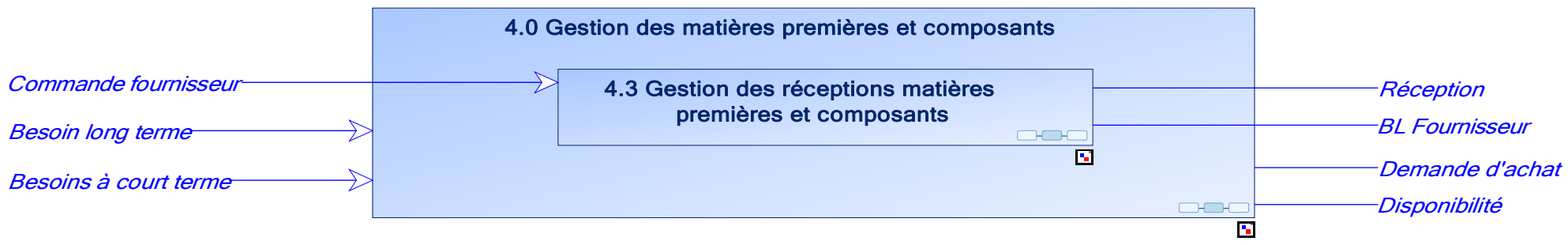

*Figure 6. Processus de réception matières dérivé du PRM*

En tant que MOE de rang 1 nous avons hérité des modèles de la MOA qui nous ont permis notamment de décrire le sous-processus 4.3 dans un diagramme de mise en œuvre de processus (*Figure 7*), au travers d'une procédure particulière dans l'organisation. Chaque procédure est placée sous la responsabilité d'un acteur (de type structure) de l'entreprise. Ce diagramme a été créé à partir de la procédure métier existante (« Réception MP, fournitures ») décrite dans le cadre du Système de Management de la Qualité de l'AIPL.

Les messages et exigences relatifs au sous-processus parent ont été intégralement récupérés au niveau des procédures. Ce diagramme a été placé dans la cellule « Model-How » du CMZ de notre SAF-eProduction. A ce niveau de décomposition, nous nous sommes posé la question : existe-t-il à l'AIPL un ou plusieurs systèmes contributeurs (COTS) susceptibles de remplir les fonctions demandées et de répondre aux exigences relatives ? La réponse étant positive, l'aspect récursif de notre démarche a impliqué la création d'une nouvelle ingénierie supportée par un nouveau CMZ pour chacun de ces sous-systèmes. Les finalités propres de chacun d'entre eux ont été formalisées dans des diagrammes d'objectifs et d'exigences particuliers qui ne doivent évidemment pas être en contradiction avec les objectifs et les exigences du système-parent auquel ils appartiennent.

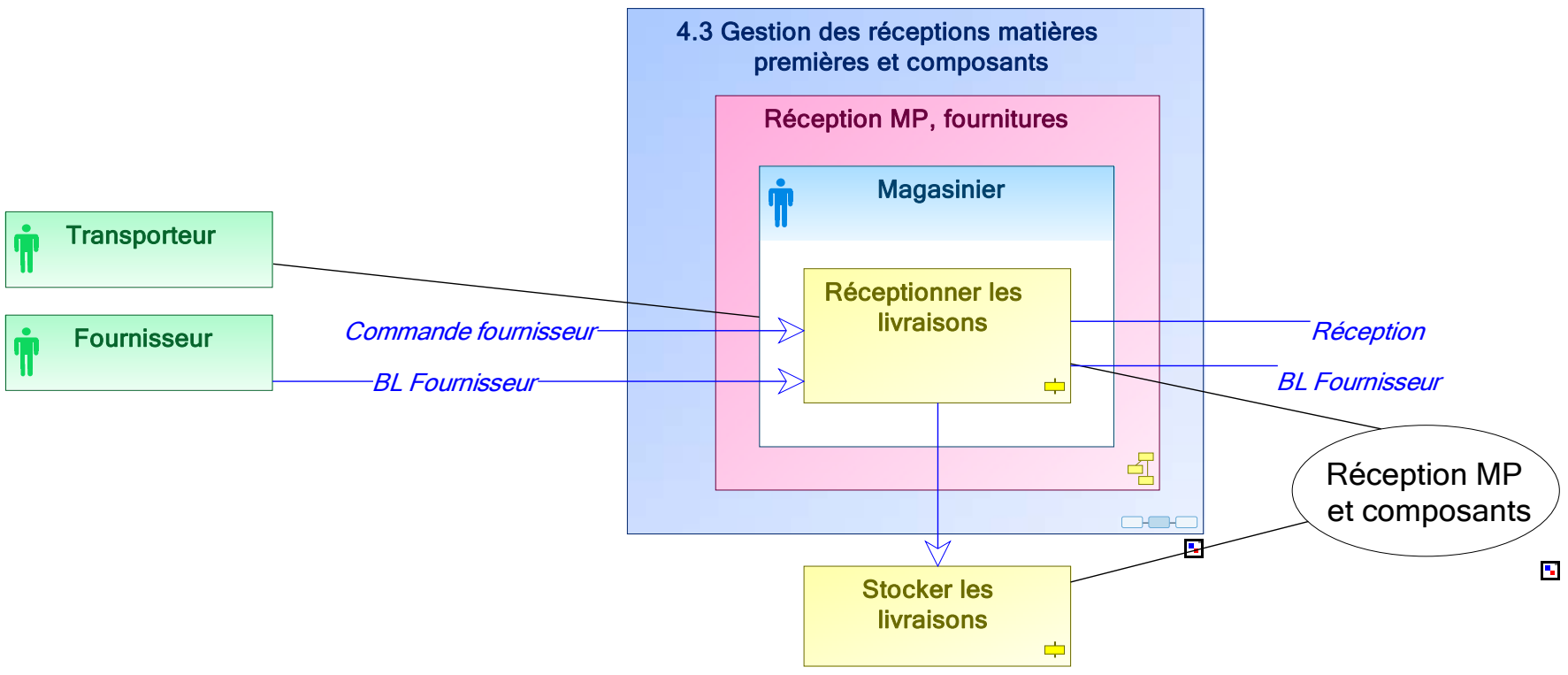

*Figure 7. Diagramme de mise en œuvre de processus*

A titre d'exemple, dans notre cas d'étude, l'exigence du système-parent SAF-eProduction « Retrouver les lots de matières premières utilisées dans un produit » implique d'une part que nos produits soient identifiés, ce qui est réalisé par un étiquetage Code Barre géré par le logiciel Akewa et que la généalogie soit enregistrée, fonction qui est supportée dans notre MES FlexNet tout au long du processus de transformation de la matière en produit fini. Si aucun COTS n'avait pu satisfaire les exigences relatives au processus, il aurait fallu créer l'ingénierie particulière d'une nouvelle application à développer ou à déployer.

Comme nous devions réaliser une ingénierie particulière, nous avons associé aux opérations de la procédure des Cas d'Utilisation au sens UML qui nous ont permis de poursuivre la modélisation d'un point de vue informatique. Les spécifications issues de la description UML du cas d'emploi « Réception MP et composants » ont alors été transmises à la MOE de rang 2 afin qu'elle puisse implémenter cette fonctionnalité de la solution sous FlexNet (création des fonctions, du logigramme représentant leur enchainement et des interfaces pour l'opérateur). Dans le même temps, chaque nouveau COTS (autonome, évolutif, hétérogène) participant à la finalité du système cible s'est inscrit dans un nouveau paquetage qui a enrichi (critère d'appartenance) le diagramme de paquetage de notre SAF-eProduction. Dans la mesure où nous avons considéré notre SI comme un SdSI [**1**], nous avons du nous focaliser sur la problématique de connectivité (à la fois sémantique, syntaxique et technique) entre COTS contributeurs.

En nous basant sur le paradigme « Système Contrôlé par le Produit », nous avons considéré le produit comme un COTS particulier formalisé par son propre paquetage. Nous avons analysé les modèles logiques des applications impliquées dans le processus de réception matières, et proposé une formalisation partielle des concepts qui leurs sont communs pour les associer ensuite à un modèle sémantique, propre à chaque COTS, placé dans la cellule « Conceptual-What » du CMZ. À partir des modèles sémantiques, nous avons proposé une correspondance sémantique des concepts respectivement de SAGE X3 (*Figure 9*) vers le Produit (*Figure 8*) et du Produit vers FlexNet. Ces correspondances nous ont permis de filtrer sur les modèles logiques correspondants, les champs implémentant les concepts modélisés et leurs interrelations.

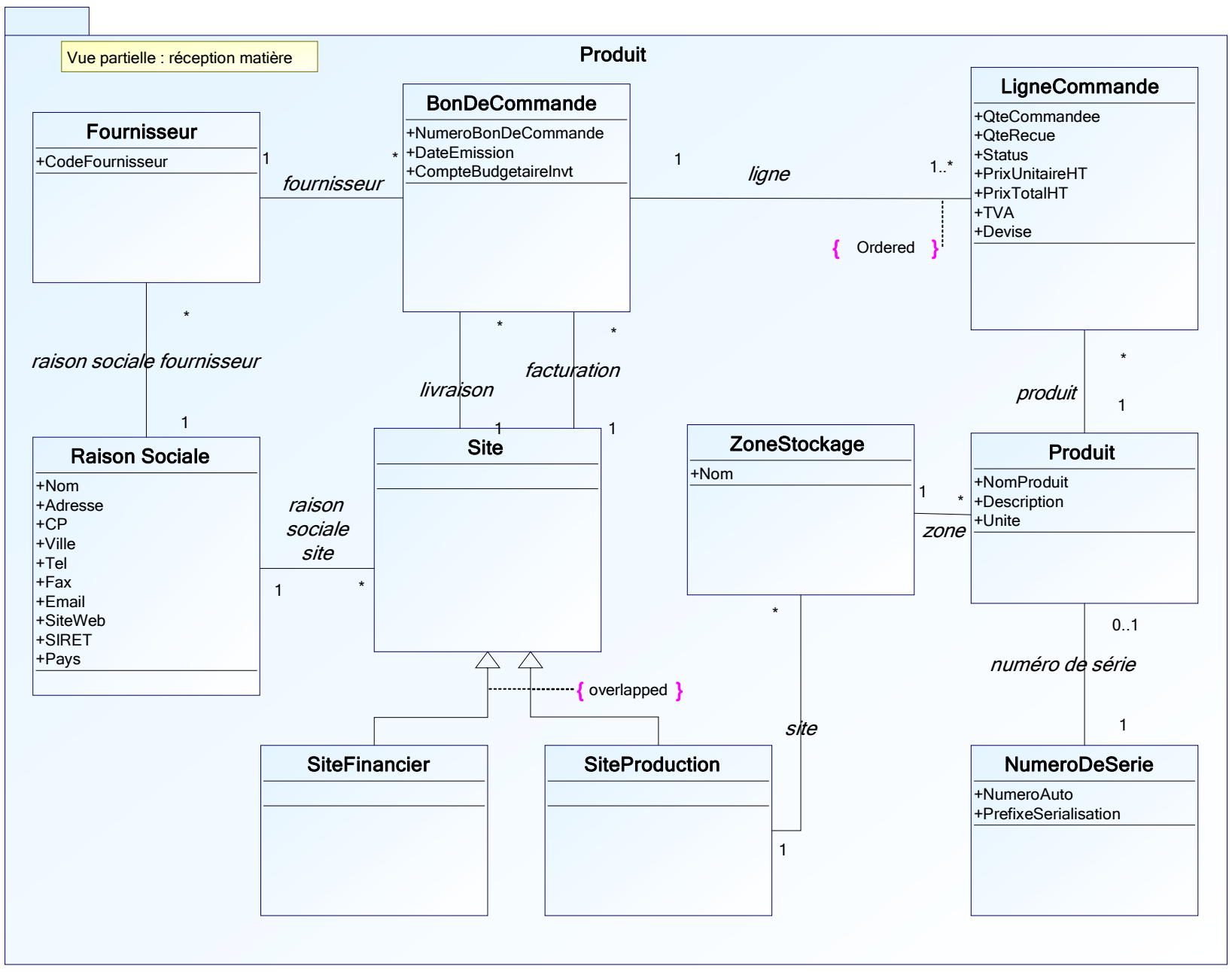

*Figure 8. Conceptualisation partielle du modèle sémantique du produit relatif au SAF « système de traçabilité du produit » à l'AIPL focalisé sur la réception matières*

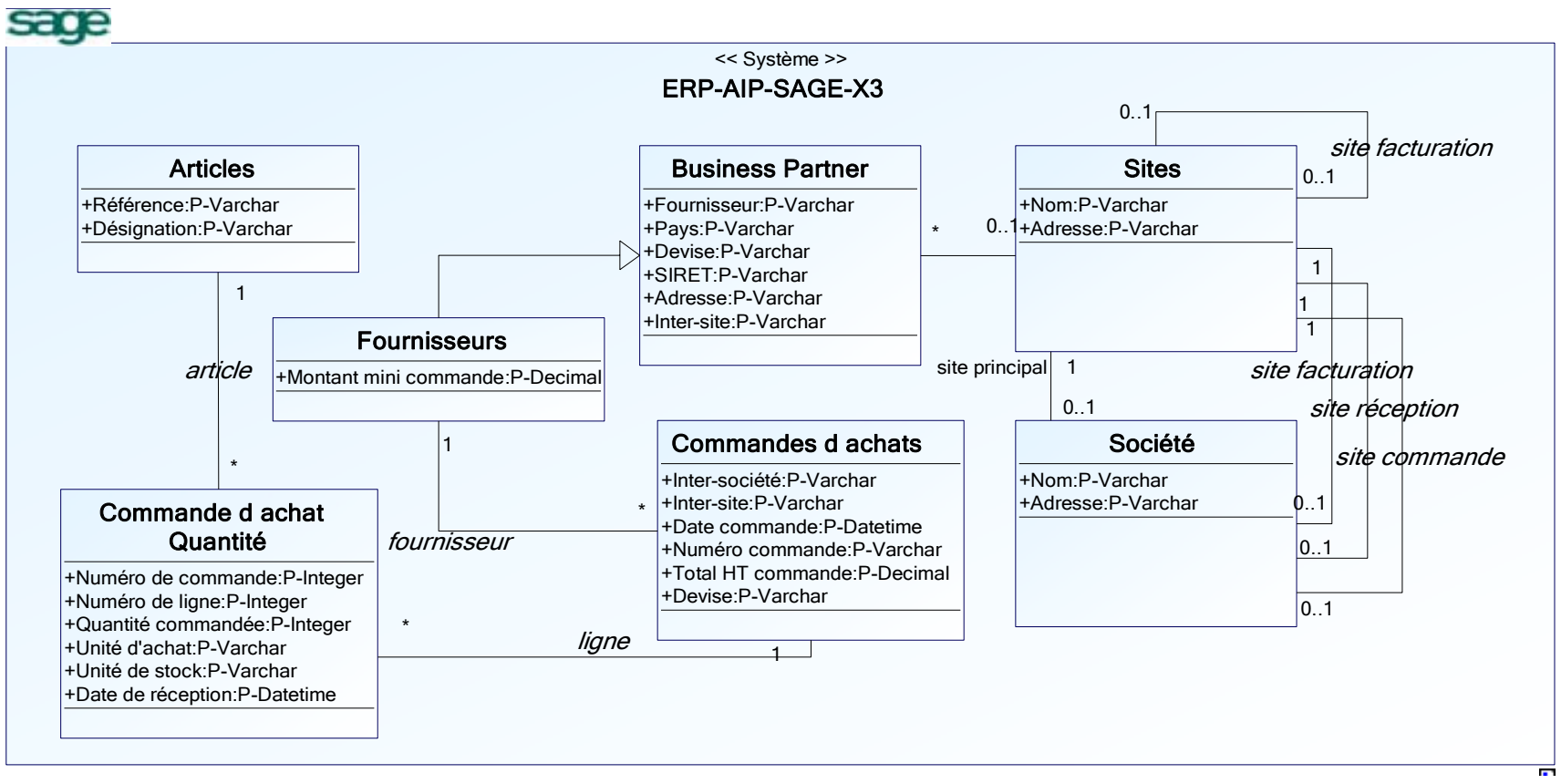

*Figure 9. Conceptualisation partielle du modèle logique interne à SAGE X3*

C'est à partir de ces diagrammes que la MOE de rang 2 a pu en déduire la structure syntaxique des flux d'informations XML échangés (XSD, DTD) et générer les fichiers de transformations au format XSLT à l'aide d'un logiciel de mapping.

Guidée par l'ensemble de ces modèles, la configuration des COTS (paramétrage des progiciels, des masques d'impression, définition des répertoires d'échange, fréquences de synchronisation, paramétrage des services etc.) est devenue plus rapide et a permis de satisfaire aux exigences techniques de connectivité définies dans le diagramme d'Objectifs et d'Exigences du SAF-eProduction.

## 5 Retour d'expérience et conclusion

Les travaux de recherche [**1**] se sont appuyés sur l'absence de capitalisation de l'ingénierie de l'interopérabilité. Voilà pourquoi, il a été choisi de se baser sur un cadre de modélisation capable de décrire un système complexe composé de nombreux systèmes constitutifs suivant des niveaux d'observations et d'abstractions variés : le CMZ. Ainsi, ce cadre agit comme un filtre structuré et multi points de vue dédié à notre Ingénierie Système. De la sorte, en se basant sur la récursivité des systèmes, il a postulé que leur ingénierie pouvait être aussi récursive sachant que certains objets de modélisation pouvaient être partagés entre l'ensemble des modèles des systèmes constitutifs et du système bénéficiaire : les invariants de modélisation. À partir du prototype permettant de capitaliser des modèles autour d'un référentiel commun assurant leur cohérence, nous avons pu démontrer au travers de notre cas d'étude à dimension suffisante que sa méthodologie pouvait être un support à l'ingénierie de système de systèmes d'information. Nous avons aussi remarqué que l'appropriation de la méthode a été relativement rapide par les modélisateurs. Il a fallu environ un mois temps plein pour assimiler la méthode et dégager les spécifications par la MOA et la MOE de rang 1. Dans notre cas la MOA et la MOE de rang 1 étaient représentées par les mêmes personnes. Les modélisateurs ont fait remarquer que la méthode leur a permis de les guider de manière rigoureuse dans le respect des exigences initiales et la cohérence des objets de modélisation. Le passage des exigences implicites aux exigences explicitement représentées dans le diagramme des exigences est une étape clef et fondamentale qui mérite d'y consacrer du temps. Une fois le diagramme des exigences approuvé par le donneur d'ordres, il a fallu environ 15 jours pour réaliser l'ingénierie de la réception matières. Il a été noté aussi que le travail de l'implémentation a grandement été facilité par le cadre rigoureux qu'impose le CMZ. Nous avons remarqué aussi que plus les modélisateurs progressaient dans leur modélisation, plus ils étaient efficaces et rapides dans leur tâche de modélisation. D'autre part, nous avons constaté que cette proposition méthodologie nous a guidés dans la spécification de nos besoins en termes de support et de formation auprès des éditeurs de COTS, ce qui nous a permis d'internaliser et donc de capitaliser leur ingénierie de connectivité technique. Cette proposition de découpage du projet « *eProduction* » en sous-projets (dans notre cas le *SAF-traçabilité_du_produit*) qui peuvent à leur tour être décomposés récursivement en sous-projets (dans notre cas le *SAF-traçabilité_du_produit_en réception_matières*), nous a permis de garantir l'atteinte des objectifs intermédiaires et le maintient en cohérence des ingénieries des éléments constitutifs hétérogènes (COTS). Depuis, cette méthode a été employée dans le cadre d'un autre sous-projet et a permis de dégager les premières spécifications tout en respectant les exigences et objectifs du projet « eFormation en eProduction ».

## 6 Glossaire

| | | | |
|---|---|---|---|
| **AIPL** | Atelier Inter-établissements de Productique Lorrain. | **AFIS** | Association Française d'Ingénierie Systèmes |
| **B2M** | Business to Manufacturing | **B2B** | Business to Business |
| **CMZ** | cadre de modélisation de Zachman | **BPMN** | Business Process Modelling Notation |
| **CRM** | Customer Relationship Management | **CIM** | Computer Integrated Manufacturing |
| **ERP** | Enterprise Resource Planning | **COTS** | Components Off The Shelf |
| **ISBM** | Ingénierie Système Basée sur les Modèles | **DIMEG** | Dipartimento di Ingegneria Meccanica e Gestionale – Politecnico di Bari |
| **MOA** | Maîtrise d'Ouvrage | **IS** | Ingénierie Système |
| **PRM** | **Purdue Reference Model** | **MES** | Manufacturing Enterprise System |
| **SAF** | **Système à Faire** | **MOE** | Maîtrise d'Œuvre |
| **SCP** | **Système Contrôlé par le produit** | **SdS** | Système de Systèmes |
| **SPF** | **Système pour Faire** | **SdSI** | Système de Systèmes d'Information |

*Table 1. Glossaire.*

## 7 Références

1. J. P. Auzelle, "Proposition d'un cadre de modélisation multi-échelles d'un système d'information en entreprise centré sur le produit," Thèse de Doctorat de l'Université Henri Poincaré, 2009.

2. J. P. Meinadier, Découvrir et comprendre l'ingénierie système (version 2008), Rédigé par le Groupe de Travail Ingénierie Système de l'Association Française d'Ingénierie Système (AFIS), 2006.

3. ISO/IEC-15288, "System life cycle processes and its guide iso," ISO TC 184/SC7/JTC1, Geneva, Switzerland, 2002.

4. S. Y. Nof, F. G. Filip, A. Molina, L. Monostori and C. E. Pereira, Advances in e-manufacturing, e-logistics, and e-service systems, Proceedings of the 17th World Congress of the IFAC, The International Federation of Automatic Control, IFAC, 2008.
5. P. Baptiste, A. Bernard, J.-P. Bourrières, P. Lopez, G. Morel, H. Pierreval and M.-C. Portmann, "Comité d'experts productique : Prospectives de recherche," 2007.
6. D. Carney, D. Fisher and P. Patrick, "Topics in interoperability: System-of-systems evolution," Carnegie-Mellon Univ Pittsburgh Pa Software Engineering, Inst, Defense Technical Information Center, 2005.
7. P. Oberndorf, "Cots and open systems," SEI Monographs on the Use of Commercial Software in Government Systems. Software Engineering Institute. Carnegie Mellon University, Pittsburgh, Pennsylvania, vol. 15213, Software Engineering Institute, Carnegie Mellon University, 1998.
8. V. Chapurlat, "Vérification et validation de modèles de systèmes complexe : Application à la modélisation d'entreprise.," Habilitation à Diriger les Recherches de Université de Montpellier 2, 2007.
9. M. W. Maier, Architecting principles for systems-of-system, Systems Engineering 1 (1998), no. 4, 267-284.
10. D. DeLaurentis, Understanding transportation as system-of-systems design problem, 43 rd AIAA Aerospace Sciences Meeting and Exhibit, 2005.
11. J. Boardman and B. Sauser, System of systems–the meaning of of, System of Systems Engineering, 2006 IEEE/SMC International Conference on (2006), 4-10.
12. D. Luzeaux and J. R. Ruault, Ingénierie des systèmes de systèmes - concepts et illustrations pratiques, vol. Vol.1, Hermes Science, 2008.
13. L. Ferrarini, A. Kalogeras, A. Lüder and al., "Deliverable 6.1, pabadis’promise, next generation control devices," PABADIS based Product Oriented Manufacturing Systems for Reconfigurable Enterprises, 2006, p. 77.
14. H. Panetto, "Meta-modèles et modèles pour l’intégration et l’interopérabilité des applications d’entreprises de production," Habilitation à Diriger des Recherches de l'Université Henri Poincaré, 2006.
15. A. W. Wymore, Mbse : Model-based systems engineering, CRC Press, Inc.
16. J. A. Zachman, A framework for information systems architecture, IBM Systems Journal 26 (1987), no. 3, 276-292.
17. J. F. Sowa and J. A. Zachman, Extending and formalizing the framework for information systems architecture, IBM Systems Journal 31 (1992), no. 3, 590-616.
18. A. W. Wymore, Model-based systems engineering, CRC Press, Inc. Boca Raton, FL, USA, 1993.
19. J.-M. Favre, J. Estublier and M. Blay-Fornarino, L'ingénierie dirigée par les modèles : Au delà du mda,traité ic2, série informatique et systèmes d'information, Lavoisier, 2006.
20. S. Baïna, "Interopérabilité dirigée par les modèles : Une approche orientée produit pour l'intéropérabilité des systèmes d'entreprise," Université Henri Poincaré, 2006.
21. A. Tursi, H. Panetto, G. Morel and M. Dassisti, Ontology-based products information interoperability in networked manufacturing enterprises, Proceedings of the IFAC CEA'2007 conference on Cost Effective Automation in Networked Product Development and Manufacturing, Elsevier - IFAC Papersonline, 2007.
22. MESA-international, "Mesa-international, http://www.Mesa.Org/."
23. PRM, "Purdue reference model for cim, http://www.Pera.Net/pera/purduereferencemodel/referencemodel.Html."